\documentclass[global,twocolumn]{svjour}
\usepackage{latexsym}
\usepackage{graphics}
\journalname{Applied Physics B}

\begin{document}

\title{A slow gravity compensated Atom Laser}

\author{G.~Kleine~B\"uning\inst{1}\thanks{\emph{e-mail:} kleinebuening@iqo.uni-hannover.de} \and J.~Will\inst{1} \and W.~Ertmer\inst{1} \and C.~Klempt\inst{1} \and J.~Arlt\inst{2}
 }

\institute{Institut f\"ur Quantenoptik, Leibniz Universit\"at Hannover, Welfengarten 1, D-30167 Hannover 
\and QUANTOP, Danish National Research Foundation Center for Quantum Optics, Department of Physics and Astronomy, Aarhus University, Ny Munkegade 120, DK-8000 Aarhus C}

\date{Received: date / Revised version: date}

\maketitle

\begin{abstract}
We report on a slow guided atom laser beam out\-coupled from a Bose-Einstein condensate of $^{87}$Rb atoms in a hybrid trap. The accelera\-tion of the atom laser beam can be controlled by compensating the gravitational acceleration  and we reach residual accelerations as low as $0.0027$~g. The outcoupling mechanism allows for the production of a constant flux of $4.5 \times 10^6$ atoms per second and due to transverse guiding we obtain an upper limit for the mean beam width of $4.6~\mu$m. The transverse velocity spread is only $0.2$~mm/s and thus an upper limit for the beam quality para\-meter is M$^2=2.5$.
We demonstrate the potential of the long interrogation times available with this atom laser beam by measuring the trap frequency in a single measurement. The small beam width together with the long evolution and inter\-rogation time makes this atom laser beam a promising tool for continuous interferometric measurements.
\end{abstract}

\begin{figure*}
	\centering
		\resizebox{1.0\textwidth}{!}{\includegraphics{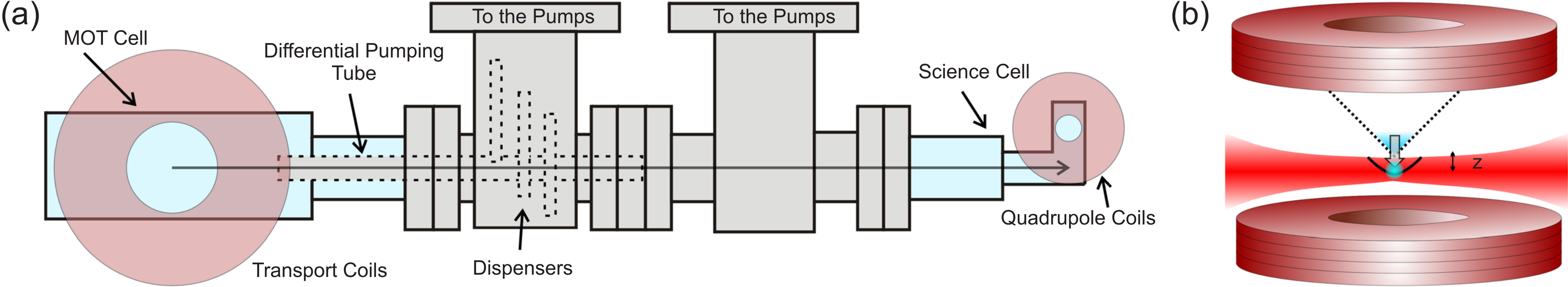}}
		\caption{Outline of the experimental apparatus. The vacuum system with the two glass cells is shown in (a). In the large MOT cell, the atoms are captured from the background vapor. Subsequently, they are trapped magnetically and transported into the science cell with movable coils. This cell is designed in a L-shape to grant optical access along three axes. The evaporative cooling takes place in a hybrid trap (b). The trapping potential is provided by a  dipole beam focused slightly underneath the center of a magnetic quadrupole field.}
	\label{fig:vacuum}
\end{figure*}

\section{Introduction}
\label{intro}

The conceptual similarities between the coherence of laser light and the coherence of Bose-Einstein condensed samples has been a driving force for the field of cold quantum gases. Similarly to the laser, coherent matter waves hold the promise of improving precision measurements and fundamental tests of quantum physics.

However, there is a striking difference between the laser and cold quantum gases that still has to be overcome. Current exponents with cold quantum gases rely on the technique of forced evaporative cooling to achieve the desired temperatures and densities. Since this cooling technique depends on collisions between the particles, conservative magnetic or optical potentials are used to confine the samples. While many similarities still apply in this trapped case, the confinement impedes the production of a bright coherent matter wave output.

It was realized soon after the production of the first Bose-Einstein condensates (BEC), that the mechanism for forced evaporation could also be used as an outcoupler for matter wave packets~\cite{Mewes1997}. This technique also allowed for the production of long pulses~\cite{Bloch1999,Gerbier2001}, limited only by the size of the initial BEC. Due to the small momentum of the outcoupled atoms, this technique is however limited by the inherent interaction with the remaining BEC fraction~\cite{Riou2006}. In a different approach Raman transitions were hence used to impart a larger momentum to the outcoupled atoms~\cite{Hagley1999,Robins2006} and it was shown that these outcouplers lead to a strongly improved output beam quality~\cite{Debs2010}.

All of these outcoupling techniques rely on the transfer of atoms to magnetically untrapped states. Yet many recent experiments use all optical or hybrid potentials to produce BEC. Hence outcoupling techniques were also developed for samples confined in dipole potentials~\cite{Cennini2003,Couvert2008} and optical lattices~\cite{Anderson1998}. 

In addition to these techniques to produce coherent matter wave beams, powerful analysis techniques have been developed to investigate their coherence properties and both first and second-order coherence were confirmed~\cite{Bloch2000,Kohl2001,Ottl2005}.

Current experimental efforts focus on methods to con\-tin\-uously replenish a BEC while simultaneously outcoupling a coherent beam. Both the replenishment  of a BEC in an optical dipole trap~\cite{Chikkatur2002} and simultaneous pumping and outcoupling~\cite{Robins2008} of limited duration were demonstrated. However completely new experimental approaches such as continuous condensation in a beam~\cite{Lahaye2004} or continuous loading of a trap~\cite{Aghajani2009} may be necessary to achieve this ambitious goal.

The similarity between a laser beam and the atom laser points at another major experimental challenge. While the internal degrees of freedom can be controlled precisely in quantum gases, control of the external degrees of freedom of an outcoupled beam poses a larger challenge. The divergence of the beam and its mode structure have been investigated intensely~\cite{LeCoq2001,Koehl2005,Riou2006,Jeppesen2008,Debs2010,Cennini2008}. First atom optical elements were developed using inhomogeneous magnetic~\cite{Bloch2001,Kohl2001} and optical~\cite{Bongs1999,Bongs2001} potentials. Recently, guided atom lasers were realized starting from optically~\cite{Couvert2008} and magnetically confined~\cite{Guerin2006} samples.

Within our work a particularly slow atom laser beam was realized. By compensating the gravitational acceleration with an inhomogeneous magnetic field, accelerations below $0.003$~g can be achieved. Thus this atom laser allows for interrogation times of up to 500~ms primarily limited by the field of view of the detection system. These long interrogation times promise continuous interferometric measurements in this system over relatively long periods of time. Special emphasis is given to the novel features of the experimental apparatus.

\section{Experiment}
\label{exp}
The investigation of slow atom laser beams is performed in an experimental apparatus based on a hybrid trap~\cite{Lin2009}. To incorporate this trap, a two chamber vacuum system shown in Fig.~\ref{fig:vacuum} is used. The two parts are separated by a differential pumping stage and allow for efficient loading of a magneto-optical trap (MOT) from the background vapor and for long lifetimes of the atomic samples in the science cell. The transport of the atoms between the cells is realized with a movable pair of quadrupole coils~\cite{Klempt2008,Lewandowski2003}. The science cell is designed in a L-shape to provide outstanding optical access from six directions.

\subsection{MOT and magnetic transport}
\label{motandtransp}

To enable the use of large trapping beams, the MOT cell has inner dimensions of $50$~mm~$\times$~$50$~mm~$\times$~$140$~mm. Rubidium vapor is provided by commercial dispensers located approximately $20$~cm from the MOT center within a direct line-of-sight. During dispenser operation the pressure reaches  $5\times10^{-10}$~mbar in the MOT cell and $1\times10^{-11}$~mbar in the science cell.

For operating the $^{87}$Rb MOT, a laser system provides light at a wavelength around $780$~nm. The cooling light on the D2 transition $F=2 \rightarrow F'=3$ is derived from a stabilized external cavity diode laser (ECDL). The repumping light on the transition $F=1 \rightarrow F'=2$ is delivered by a second ECDL which is referenced to the cooling laser with a frequency offset of $6.8$~GHz. Both beams are amplified to a total power of $1$~W in a single tapered amplifier~\cite{wilson1998} and delivered to the experiment in a polarization maintaining optical fiber. To operate the MOT, the total available laser power of $350$~mW is divided into six beams which are then individually expanded to a diameter of $50$~mm in Galilean telescopes. 

The large beams in combination with the large glass cell enable for efficient and fast loading from the Rubidium background vapor. The loading rate is further enhanced by the use of ultraviolet light-induced atom desorption~\cite{Klempt2006} at $395$~nm which increases the atom number by a factor of 2 after 10~s loading time, resulting in $3\times10^9$~atoms. Once this atom number is reached, the MOT is turned off and the atoms are further cooled within an optical molasses phase and optically pumped to the $\mid\! F,m_F\rangle=\mid\! 2,2\rangle$ for further magnetic trapping.

To transfer the atoms to the science cell, they are transported over a distance of $60$~cm within $1.2$~s by a quadrupole magnetic field. The MOT coils are therefore mounted on a movable translation stage. For the transport, the current through these coils is increased to $45$~A. The corresponding magnetic field gradient of $B'=165$~G/cm (where $B'=\partial B / \partial z$ denotes the gradient in the strongest, vertical direction) captures and compresses the cold cloud. Once the translation stage reaches its final position at the science cell, the atoms are transferred over a distance of $4.5$~cm into a second pair of quadrupole coils, as shown in Fig.~\ref{fig:vacuum}. We obtain $2 \times 10^8$~atoms at a peak phase space density of $1 \times 10^{-7}$ which provides an excellent starting condition for further cooling in the hybrid trap.

\subsection{Hybrid trap}
\label{hybrid}
Since the typical phase space density after laser cooling is still far from quantum degeneracy, all BEC experiments to date employ evaporative cooling to reach the required temperatures. In most cases this is done either by forced radio-frequency evaporation in magnetic potentials or by lowering the trap depth in optical dipole potentials. In our experiment we combine the two approaches in a hybrid trap, profiting from the advantages of both~\cite{Lin2009}.

The magnetic confinement is provided by a simple magnetic quadrupole potential which initially allows for very efficient evaporation. Additional optical confinement is realized with a single gaussian beam with a waist of $52~\mu$m, focused below the center of the quadrupole potential. We use a single-mode, single-frequency fiber laser at a wavelength of $1064$~nm which provides a power of up to $7$~W at the position of the atomic cloud. The power is controlled with an acousto-optic modulator, whose control circuit is optimized for stability at low optical power of a few milliwatts. 

After the transfer of $2 \times 10^8$ atoms to the quadrupole trap we employ forced microwave evaporation on the $\mid\!2,2\rangle$ to $\mid\!1,1\rangle$-transition. Simultaneously, we ramp the magnetic field gradient from $B'=300$~G/cm down to $132$~G/cm. The evaporation is stopped at $6.845$~GHz which corresponds to a magnetic field of $5$~G. After this first evaporation step, $7 \times 10^7$ atoms at a phase space density of $2 \times 10^{-4}$ are obtained. If this evaporation in the quadrupole potential is continued, the lifetime begins to suffer from Majorana spin-flip losses.

At this point, the atoms are therefore transferred into the hybrid trap shown in Fig.~\ref{fig:vacuum}. This is accomplished by increasing the dipole beam power to $7$~W while lowering the magnetic gradient to
\begin{equation}\label{eq:bcomp}
   B'_{\rm{c}} = m_{\rm{Rb}} g/(m_F g_F \mu_{\rm{B}})=15 \; \rm{G/cm} \; .
\end{equation}
Here, $m_{\rm{Rb}}$ denotes the mass of $^{87}$Rb, $g$ the gravitational acceleration, $g_F$ is the Land\'e factor and $\mu_{\rm{B}}$ represents Bohr's magneton. During this process, the atoms sag under the influence of gravity and are thus transferred into the optical dipole potential. At the final value $B'_{\rm{c}}$, gravity is compensated and thus, the atoms are primarily confined by the dipole potential in radial direction.

\begin{figure}
	\centering
	\resizebox{\linewidth}{!}{	\includegraphics{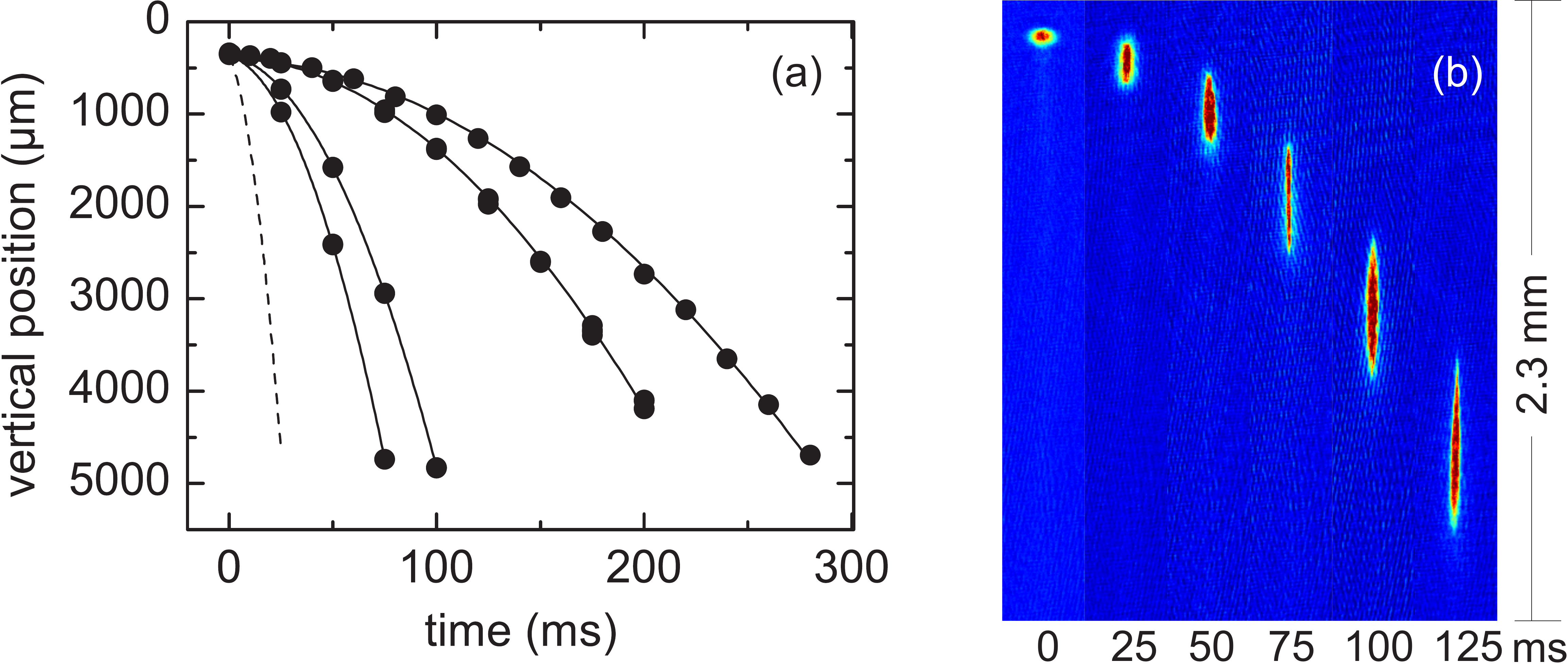}}
	\caption{(a) BEC positions after time of flight in the presence of a magnetic field gradient. Residual accelerations of $0.14$~g, $0.082$~g, $0.018$~g and $0.010$~g (from left to right) are obtained from fits to this data. The dashed line indicates free fall at an acceleration of $1$~g. For the case of $0.018$~g the original absorption images are shown (b).}
	\label{fig:variousgrads}
\end{figure}
To calibrate the exact magnetic field gradient, we measure the residual acceleration of BECs released from the hybrid trap in the presence of various field gradients. Figure~\ref{fig:variousgrads} shows an exemplary series of absorption images and the positions of the BEC as a function of the time of flight.

In the axial direction, the dipole potential hardly provides any confinement and the trap relies on the remaining magnetic field. The quadrupole potential is linear on axes through the center, but it has a curvature everywhere else. The quadrupole potential is given by
\begin{equation}\label{eq:umag}
	U(x,y,z)=m_F g_F \mu_{\rm{B}} B' \ \sqrt{\left.\left(x^2+y^2\right)\right/4+ z^2}\; ,
\end{equation}
where $x,y$ are the horizontal coordinates and $z$ is the vertical position. The horizontal oscillation frequency $\omega_x$ is obtained by calculating the curvature at a displacement $z$ below the quadrupole center:
\begin{equation}\label{eq:trapfreq}
	 \omega_{x}(z)=\sqrt{\frac{m_F g_F \mu_{\rm{B}}}{4 m_{\rm{Rb}}}\;  \frac{B'}{z}}
\end{equation}

\begin{figure}
	\centering
	\resizebox{\linewidth}{!}{	\includegraphics{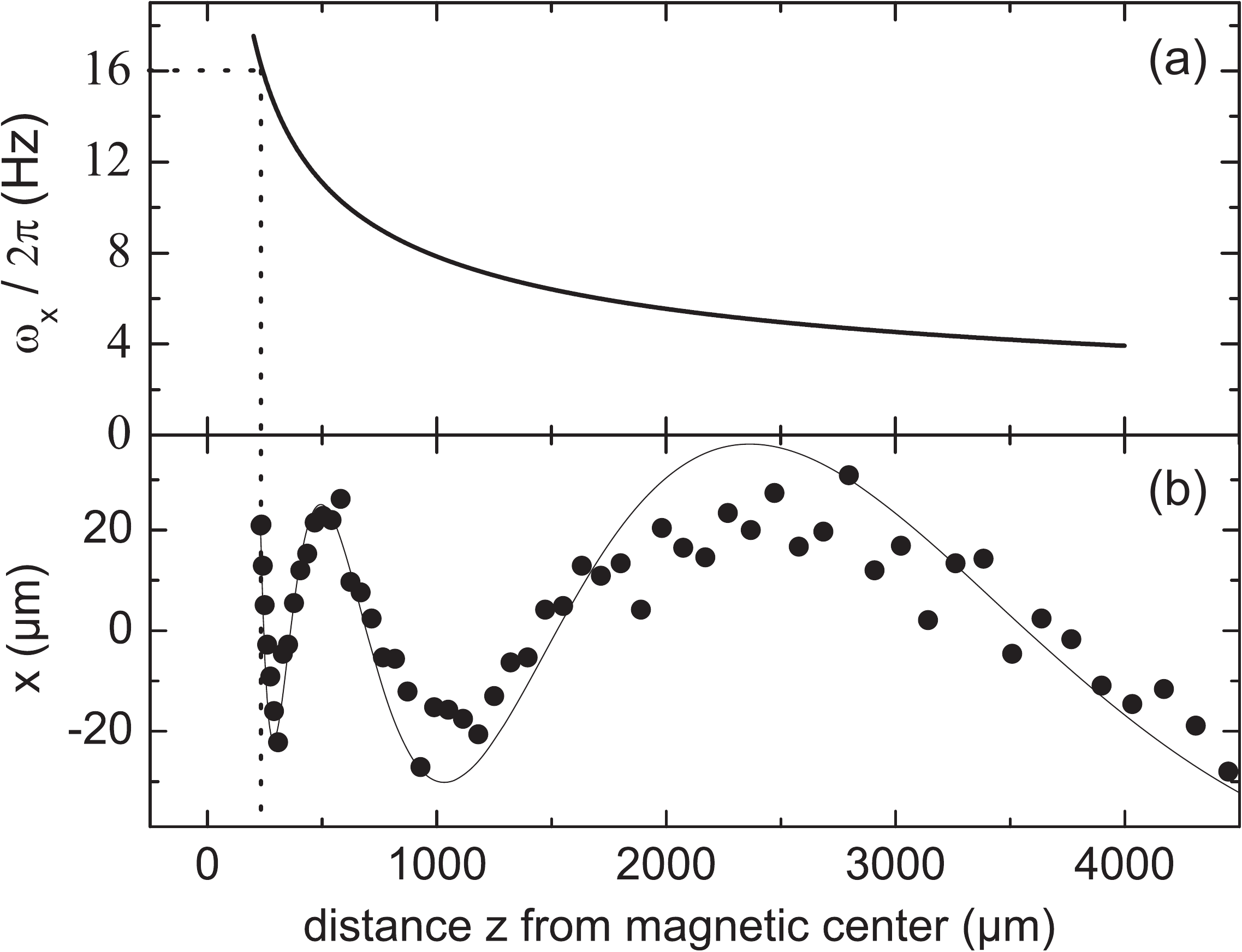}}
	\caption{(a) Horizontal oscillation frequency as a function of the vertical position below the quadrupole center. The dotted line shows the position of the hybrid trap $230~\mu$m below the quadrupole center, which results in an axial confinement of $16$~Hz. (b) Oscillating BEC falling through the magnetic field with a residual acceleration of $0.010$~g. The solid line is a solution of the equation of motion with only starting velocity and position as the free parameters.
	}
	\label{fig:becoszi}
\end{figure}
In the experiment, the dipole beam is focused $z=230~\mu$m below the center to avoid Majorana spin-flip losses and to obtain a harmonic confinement with a sufficiently high axial trap frequency of $\omega_x=2\pi \times $16~Hz. At positions below the hybrid trap the horizontal oscillation frequencies decrease as shown in Fig.~\ref{fig:becoszi}. To analyze this transverse confinement during the time of flight, we displace the BEC in the hybrid trap. It hence oscillates with an axial frequency of $16$~Hz before it is released and continues to oscillate while falling through the magnetic field. Figure~\ref{fig:becoszi} shows the horizontal position of BECs during this process. As the BEC falls into regions of smaller axial confinement, the oscillation frequency decreases and the amplitude increases. All of these features are reproduced by solving the classical equation of motion for a particle in the potential of Eq.~(\ref{eq:umag}).

During the transfer into the hybrid trap, roughly two thirds of the atoms are lost, but one order of magnitude in phase space density is gained. This fortunate behavior is caused by ongoing evaporative cooling and by an additional gain in peak phase space density due to the change from the linear quadrupole potential to the harmonic hybrid trap~\cite{Lin2009}.

With an initial atom number of $2 \times 10^7$ and a phase space density of $2 \times 10^{-3}$ in the hybrid trap, BEC is easily obtained by decreasing the dipole beam power and thus lowering the trap depth. At a laser power of $20$~mW corresponding to a trap depth of $600$~nK BECs of up to $1 \times 10^{6}$ atoms are created. 

\begin{figure}
	\centering
	\resizebox{\linewidth}{!}{	\includegraphics{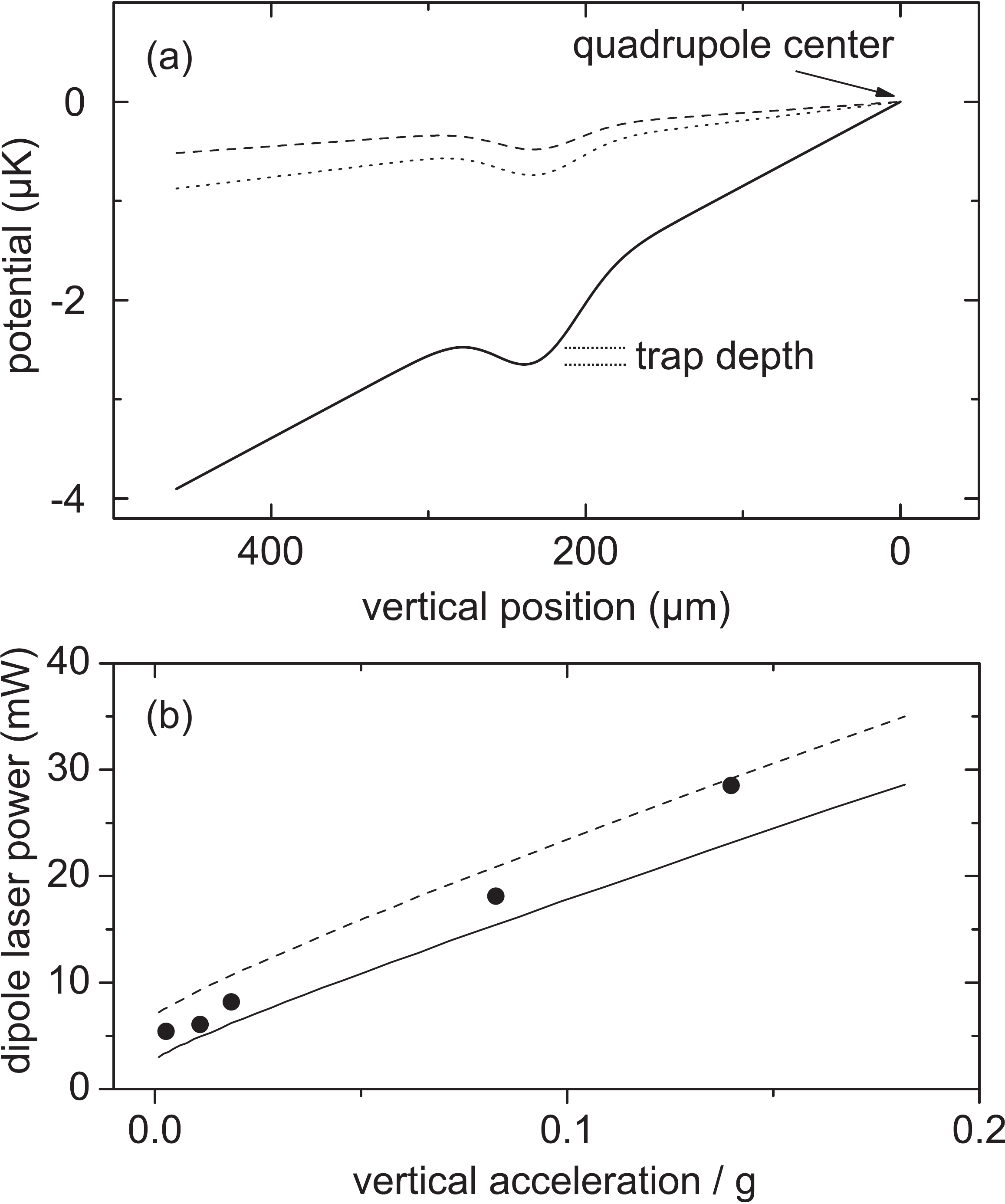}}
	\caption{(a) Potential provided by the hybrid trap for residual accelerations of 0.0010~g (dashed), 0.018~g (dotted) and 0.082~g (solid) and the corresponding minimal dipole beam power needed to support a BEC. These values are shown in (b). Reducing the dipole beam power below these values leads to outcoupling of atoms. The experimental values are compared to calculated powers for fixed trap depths of 100~nK (solid) and 250~nK (dashed).}
	\label{fig:potentials}
\end{figure}
After producing a BEC in the hybrid trap by lowering the trap depth, we can slowly ramp down the power of the dipole beam even further. This results in a cigar shaped trap with radial frequencies of typically $\omega_y=\omega_z=2\pi \times 30$~Hz and an axial frequency (dominated by the magnetic confinement) of $\omega_x=2\pi \times $16~Hz. During this process, further evaporation purifies the BEC and due to the adiabatic expansion into this very weak trap, the temperature is additionally decreased. Figure~\ref{fig:potentials} shows this potential for values of the magnetic field gradient and the experimentally determined minimal laser power needed to trap a BEC of $5 \times 10^{5}$ atoms. Thus lowering the potential saddle can serve as an outcoupling mechanism for this hybrid trap.

\section{Atom laser}
\label{atomlaser}

The distinctive features of the hybrid trap allow for the production and investigation of a gravity compensated, particularly slow atom laser beam. 

\subsection{Atom laser with gravity compensation}

To produce an atom laser beam we reduce the dipole laser power to a fixed value slightly below the threshold values presented in Fig.~\ref{fig:potentials}. The threshold value is determined by the chemical potential of the BEC which is set by the repulsive interaction of the Rubidium atoms and the number of atoms in the trap. If the trap depth falls below the chemical potential, the BEC becomes partially untrapped and the trap leaks. The leak is situated at the lowest point of the condensate's surface and the spatial size of the leak scales with the difference between trap depth and chemical potential. The atoms flow out of the leak and the repulsive interaction of the escaping atoms limits the leakage flow. For long outcoupling times, the loss of trapped atoms in the BEC decreases the chemical potential, reducing and eventually terminating the leakage. The timescale of this decrease is large in the limit of large BECs and small leakage, since the chemical potential depends on the atom number $N$ as $N^{2/5}$.

The outcoupled atom laser beam propagates within the remaining magnetic quadrupole field. This field is linear in the vertical direction and compensates gravity. In the horizontal direction it provides a symmetric quadratic confinement along the whole length of the beam.

\begin{figure}
	\centering
	\resizebox{\linewidth}{!}{	\includegraphics{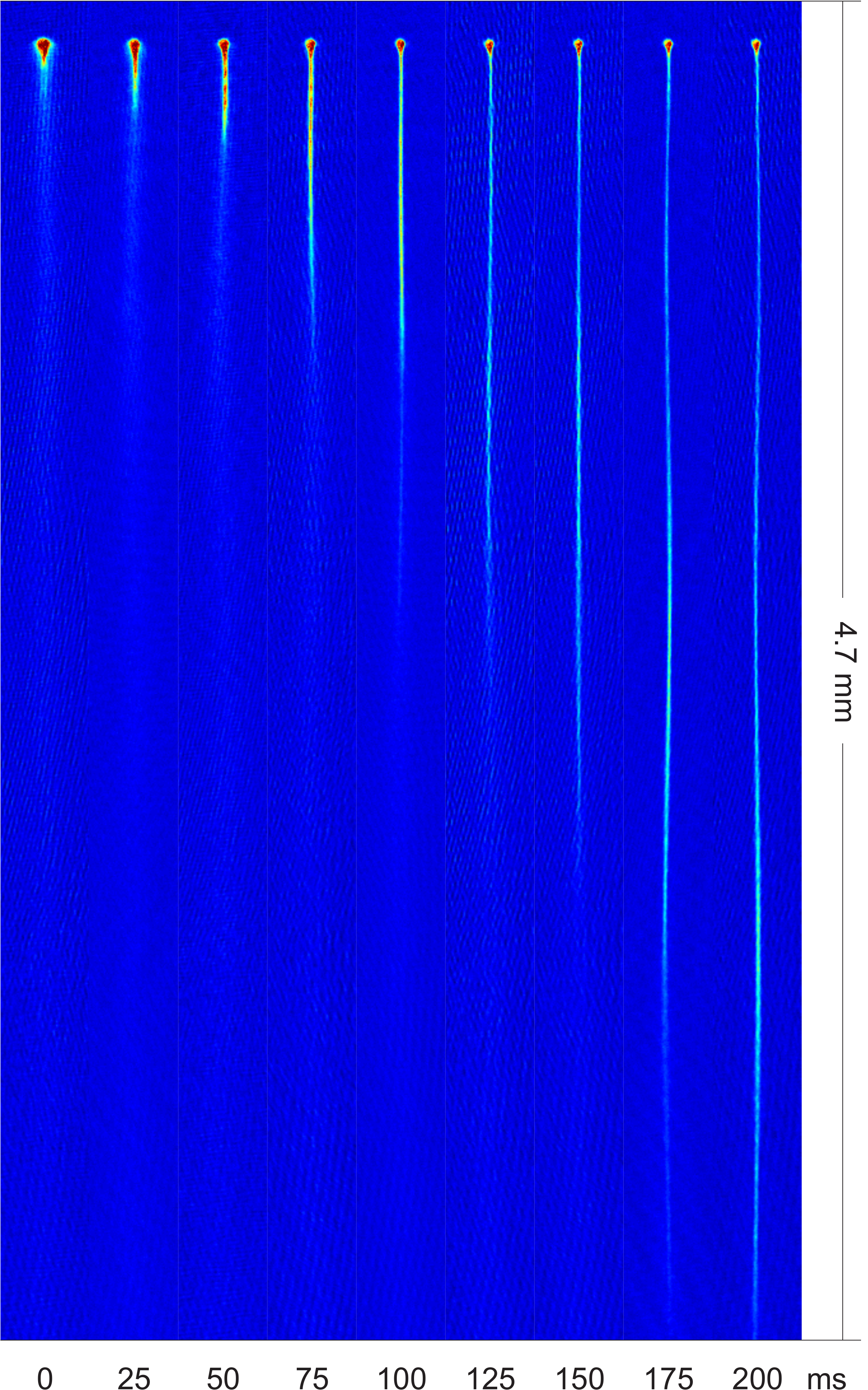}}
	\caption{Absorption images of a gravity compensated atom laser. The outcoupling time is increased from $0$~ms to $200$~ms in steps of $25$~ms.}
	\label{fig:atomlasersammlung}
\end{figure}
Figure~\ref{fig:atomlasersammlung} shows a set of absorption images of the atom laser. These images were taken for a residual acceleration of $0.018$~g and the outcoupling time was varied from zero up to $200$~ms. By choosing gradients even closer to gravity compensation, atom laser beams were observed for more then half a second at a residual acceleration of $0.0027$~g. In this case a peak velocity of only 13~mm/s after 500~ms of acceleration is reached. This demonstrates the long interrogation times available in this system, which were in fact only limited by the field of view of the detection system.

Despite of the long evolution, hardly any broadening of the width of the atom laser beam is observed within the resolution of our imaging system. This is due to the very low temperature of the BEC in the shallow hybrid trap and due to the horizontal confinement provided by the magnetic potential.

\subsection{Beam analysis}

\begin{figure}
	\centering
	\resizebox{\linewidth}{!}{	\includegraphics{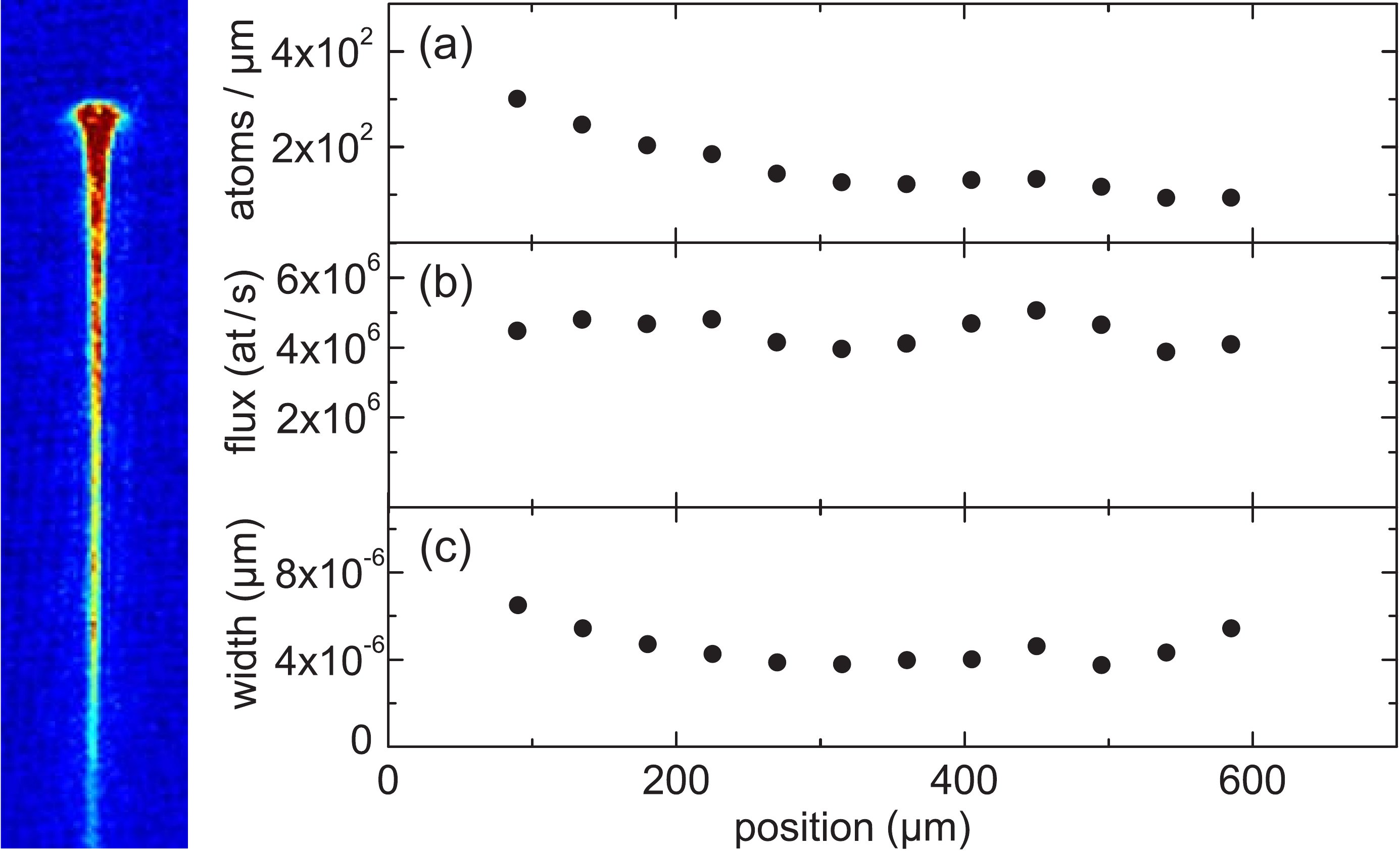}}
	\caption{Analysis of an atom laser beam coupled out of the hybrid trap at an acceleration of 0.14~g. A series of gaussian fits is performed well below the BEC over 45~$\mu$m slices in vertical direction. The atom numbers per $\mu$m (a) are used to determine the flux (b). The atom beam has a mean width of 4.6~$\mu$m (c).}
	\label{fig:beamprofile}
\end{figure}
To evaluate the beam profile quantitatively, we analyze the atom laser beam shown in Fig.~\ref{fig:beamprofile}. This beam was outcoupled for $30$~ms at an residual acceleration of $0.14$~g. At several positions within the beam the profile is averaged over $45~\mu$m slices in vertical direction. A gaussian fit is performed to obtain the width and the atom number for each of these slices. Figure~\ref{fig:beamprofile} shows the resulting atom number per $\mu$m, the calculated atomic flux and the width along the beam (limited by the resolution of our detection system). The atom number per slice decreases along the beam as expected, since it is stretched due to the remaining acceleration. The flux remains constant at approximately $4.5 \times 10^6$~atoms/s along the beam, showing that a constant outcoupling mechanism was realized. This analysis is valid as long as the size of the BEC itself diminishes only little during the outcoupling time. Note however, that the flux decreases for long outcoupling times as the chemical potential for the BEC drops (see Fig.~\ref{fig:atomlasersammlung}).

In addition to the spatial distribution shown in Fig.~\ref{fig:beamprofile}, the beam quality is determined by the transverse velocity distribution. To analyze this distribution, atom laser beams were produced and then released in absence of any external potential for times of flight up to $25$~ms. This allows for an analysis of the broadening of the beam and thus a transverse velocity spread of $0.2$~mm/s was determined.

Recent analyses of atom laser beams~\cite{Riou2006,Jeppesen2008} determine the beam quality factor $M^2$ in analogy to optical lasers. This value is defined as $M^2=(2m_{\rm{Rb}}/\hbar) \; \Delta x \Delta v_x$, where $\Delta x$ denotes the beam width at the waist and $\Delta v_x$ denotes the horizontal velocity spread. For a mean width of $4.6~\mu$m and the measured velocity spread we obtain an upper limit $M^2=2.5$ for the gravity compensated atom laser beam. This value surpasses the Heisenberg limit for an atom laser and competes with radio frequency outcoupling mechanisms~\cite{Jeppesen2008,Riou2006}.

\subsection{All-optical atom laser}

To compare the performance of the gravity compensated atom laser with an uncompensated freely propagating atom laser, an all-optical atom laser was realized~\cite{Cennini2003,Couvert2008}. The axial magnetic confinement in the hybrid trap is substituted by a second dipole laser beam. This beam crosses the first beam at an angle of $18^\circ$ in the horizontal plane. In this configuration a power of $200$~mW focused to a waist of 70~$\mu$m in the additional beam provides an axial harmonic confinement with an oscillation frequency of $25$~Hz. This confinement is sufficient to turn off the quadrupole trap completely and maintain the cold sample in a purely optical trap. By lowering the power of the first dipole beam BECs with $5 \times 10^5$ atoms are also obtained in this configuration. The final power needed to support the atoms against gravity is however considerably higher (150~mW) than in the case of the hybrid trap, which results in a more elongated trapping geometry. 

\begin{figure}
	\centering
	\resizebox{\linewidth}{!}{	\includegraphics{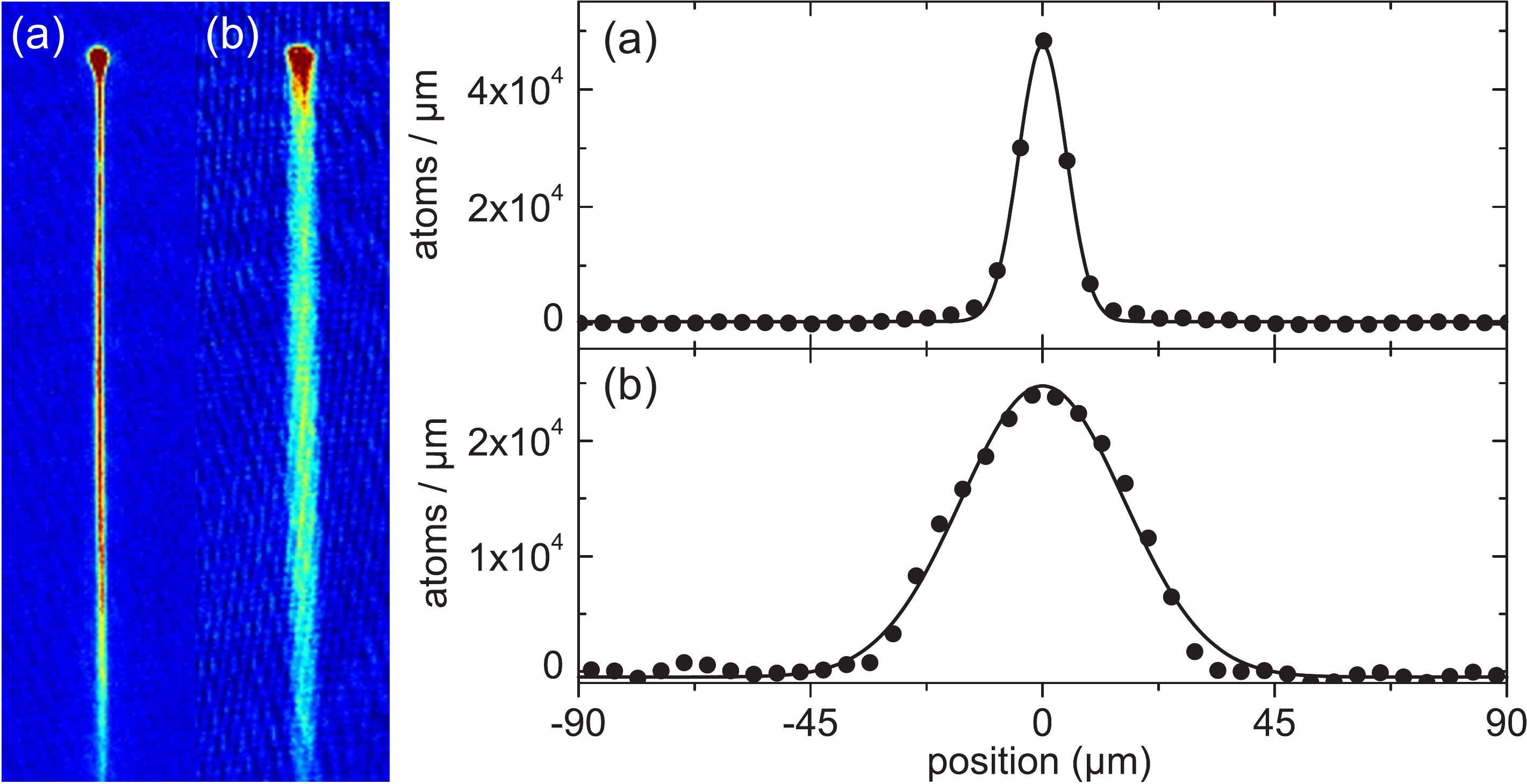}}
	\caption{Comparison of an atom laser coupled out of the hybrid trap at a residual acceleration of $0.018$~g (a) and an atom laser from a crossed dipole trap (b). The images have a vertical length of $1.3$~mm. Widths of $4.6~\mu$m and $16~\mu$m are determined with gaussian fits to the laser profiles respectively (excluding the region of the BEC).} 
	\label{fig:lasercompare}
\end{figure}
Analogously to the hybrid configuration, outcoupling into an atom laser is achieved by reducing the laser power. Figure~\ref{fig:lasercompare} shows a comparison of two atom laser beams of the same length, coupled out of the hybrid trap, and the all optical crossed dipole trap. Both are well collimated, but the gravity compensated atom laser beam has a much smaller width and thus a higher density. Moreover the quadrupole confinement is symmetric with respect to the vertical axis and thus the atom laser is also symmetric. The outcoupling of the crossed dipole trap instead depends strongly on the exact geometry of the two dipole beams, typically resulting in asymmetric beam profiles.
The outcoupling threshold is a few times higher which can be explained by the higher kinetic energy in the condensate due to the tighter confinement.

This comparison shows the superior quality of the gravity compensated atom laser beam, due to its transverse confinement in the quadrupole potential.

\subsection{Single shot measurement of the trap frequency}

An example of the beneficial long interrogation time of the atom laser is provided by a determination of the transverse oscillation frequency in a single measurement. 

\begin{figure}
	\centering
	\resizebox{\linewidth}{!}{	\includegraphics{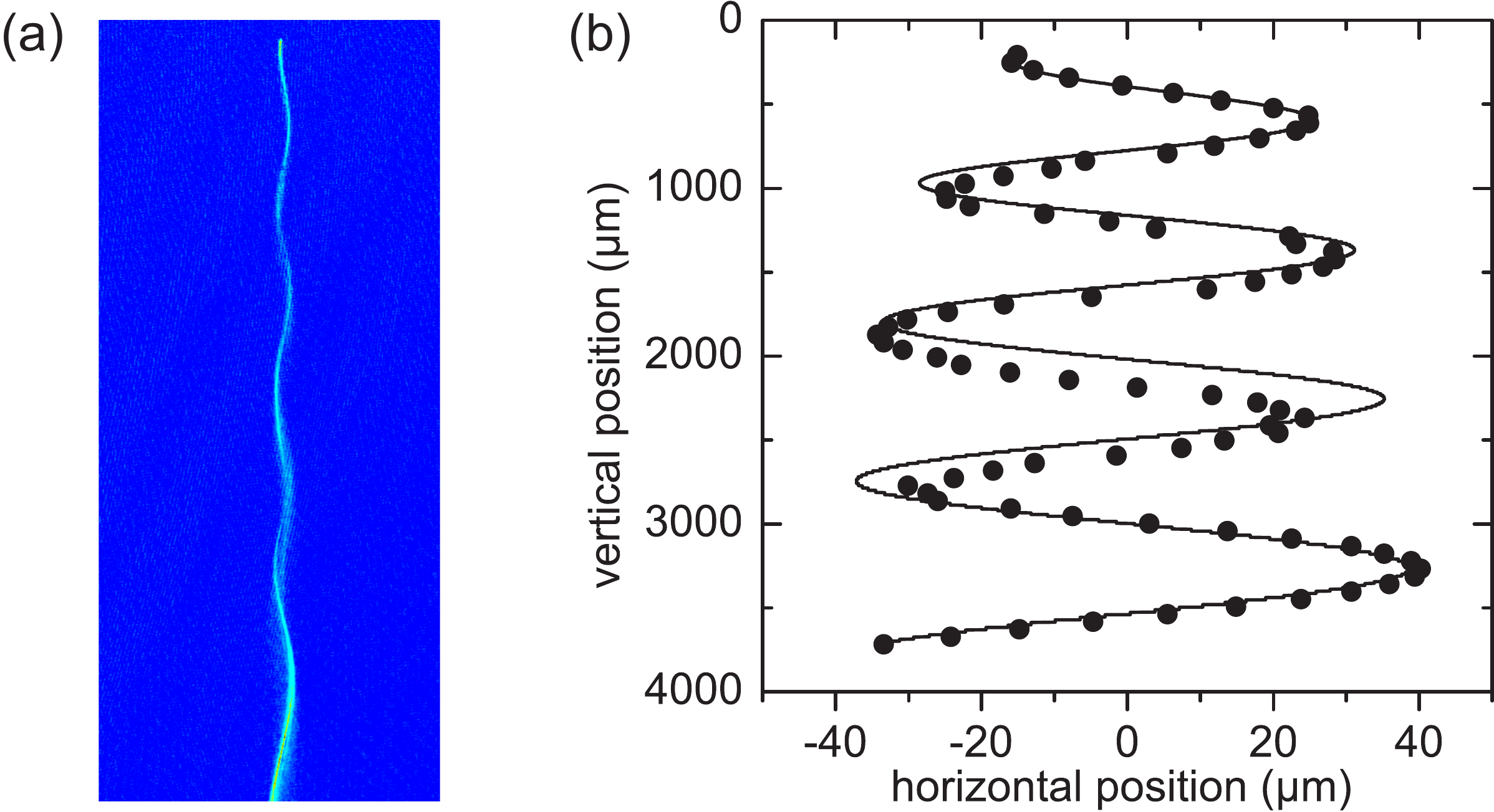}}
	\caption{(a) Absorption image of an oscillating atom laser beam at a residual acceleration of 0.0027~g. (b) Position of the atom laser beam. The solid line is a solution of the equation of motion for Eq.~(\ref{eq:umag}) taking the continuous oscillation of the source BEC into account.}
	\label{fig:laseroszi} 
\end{figure}
Figure~{\ref{fig:laseroszi} shows an absorption image of an oscillating atom laser beam and its horizontal and vertical position profile. If the BEC is set in motion in the hybrid trap before initiating the outcoupling, the beam continues its oscillations due to the horizontal confinement below the trap center. In Fig.~{\ref{fig:laseroszi} the magnetic gradient was tuned very close to gravity compensation such that the residual acceleration was only $0.0027$~g. Due to the long interrogation time, four periods of the oscillation can be observed in a single image. This presents a substantial improvement over Fig.~\ref{fig:becoszi}, requiring many individual measurements. Analogously to the discussion of Fig.~\ref{fig:becoszi}, the solution of the equation of motion for the potential Eq.~(\ref{eq:umag}) fits the data very well. For this calculation the continuous oscillation of the source BEC with a frequency of 16~Hz has been taken into account. The small discrepancy observed in the third oscillation period is explained by additional vertical oscillations of the BEC in the trap. Hence the atom laser beam can be used to measure the trap frequency with a single absorption image.

\section{Conclusion}
\label{con}
We have realized a slow guided atom laser beam outcoupled from a Bose-Einstein condensate of $^{87}$Rb atoms in a hybrid trap. The hybrid trap uses the combination of a magnetic quadrupole and an optical dipole potential to confine the atoms. The acceleration of the atom laser beam can be controlled by compensating the gravitational acceleration with the magnetic quadrupole field and allows us to reach residual accelerations as low as $0.0027$~g. In this case the atom laser beam can be monitored for 500~ms (only limited by the field of view of our imaging system), and even the highest velocity at the tip of the beam is only 13~mm/s.

An analysis of the beam shows that the outcoupling mechanism allows for the production of a constant flux of $4.5 \times 10^6$ atoms per second, while the depletion of the BEC is small. Since the magnetic quadrupole field also acts as a transverse guide, this beam is well collimated and we obtain a mean beam width of $4.6~\mu$m, far superior to an atom laser beam coupled out of an all optical potential. A time-of-flight analysis shows that the beam has a transverse velocity spread of only 0.2~mm/s and thus an upper limit for the beam quality parameter M$^2=2.5$ is obtained, comparable with the results for radio frequency outcoupling mechanisms.

Finally, the benefit of the long interrogation time available with this slow beam is demonstrated. A single image of the beam oscillating within the quadrupole potential can be used to determine the trap frequency, simplifying repeated measurements with individual BECs. The small beam width together with the long evolution and interrogation time makes this atom laser beam a promising tool for continuous interferometric measurements.

\begin{acknowledgement}

We acknowledge support from the Centre for Quantum Engineering and Space-Time Research QUEST and from
the Deutsche Forschungsgemeinschaft (European Graduate College Quantum Interference and Applications)
\end{acknowledgement}

 \bibliographystyle{unsrt}
 \bibliography{atomlaser}

\end{document}